# Multivalent ion-mediated nucleic acid helix-helix interactions: RNA versus DNA

Yuan-Yan Wu[1], Zhong-Liang Zhang[1], Jin-Si Zhang[1], Xiao-Long Zhu[2] and Zhi-Jie Tan[1*]

[1]*Department of Physics and Key Laboratory of Artificial Micro & Nano-structures of Ministry of Education, School of Physics and Technology, Wuhan University, Wuhan 430072, China*

[2]*Department of Physics, School of Physics & Information Engineering, Jianghan University, Wuhan 430056, China*

## Abstract

Ion-mediated interaction is critical to the structure and stability of nucleic acids. Recent experiments suggest that the multivalent ion-induced aggregation of double-stranded (ds) RNAs and DNAs may strongly depend on the topological nature of helices, while there is still lack of an understanding on the relevant ion-mediated interactions at atomistic level. In this work, we have directly calculated the potentials of mean force (PMF) between two dsRNAs and between two dsDNAs in $Co(NH_3)_6^{3+}$ (Co-Hex) solutions by the atomistic molecular dynamics simulations. Our calculations show that at low [Co-Hex], the PMFs between B-DNAs and between A-RNAs are both (strongly) repulsive. However, at high [Co-Hex], the PMF between B-DNAs is strongly attractive, while those between A-RNAs and between A-DNAs are still (weakly) repulsive. The microscopic analyses show that for A-form helices, Co-Hex would become "internal binding" into the deep major groove and consequently cannot form the evident ion-bridge between adjacent helices, while for B-form helices without deep grooves, Co-Hex would exhibit "external binding" to strongly bridge adjacent helices. In addition, our further calculations show that, the PMF between A-RNAs could become strongly attractive either at very high [Co-Hex] or when the bottom of deep major groove is fixed with a layer of water.

*To whom correspondence should be addressed: zjtan@whu.edu.cn

# 1. Introduction

Nucleic acids are highly charged polyanions (1). The structural folding of nucleic acids into compact native structure generally experiences strong Coulomb repulsive force, while metal ions in solution would bind to nucleic acid surface to reduce the Coulomb repulsion during folding (2-11). Therefore, metal ions play a critical role in the folding thermodynamics and kinetics of nucleic acids (2-11).

The double-stranded (ds) helix is a fundamental segment in nucleic acid structures (1). The ion-mediated interaction between two ds helices would provide the energetics for RNA structural collapse and DNA condensation (1-11). For dsDNAs, the existing experiments suggest that monovalent ions can only mediate an inter-helix repulsion, while multivalent ions can induce an attractive force and consequently cause DNA condensation (2,4-6,12-19). Such multivalent ion-mediated effective attraction has also been proposed to cause the condensation of other like-charged polyelectrolytes (20-25).

Due to the similarity between dsDNA and dsRNA in charge density (1), it is natural to assume that the ion-mediated force between dsRNAs would be similar to that between dsDNAs, and short dsDNAs have been used instead of dsRNAs to probe the ion-dependent structural assembly of nucleic acid helices (26-29) despite their different helix structures. However, the recent UV adsorption experiments have shown that in 20 mM $Na^+$ buffer, ~4mM Cobalt hexamine ($Co(NH_3)_6^{3+}$, i.e., Co-Hex) ions can cause the aggregation of short dsDNAs, but cannot lead to the aggregation of short dsRNAs (30,31). The remarkable finding is beyond the expectation and has been proposed to be attributed to the binding of Co-Hex ions into deep major groove of RNAs (30-32). However, there is still lack of a direct illustration on the linkage between ion-mediated interaction and the relevant ion-binding structure, corresponding to the experimental ionic conditions (30-32). In this work, we will directly calculate the potentials of mean force between two dsRNAs as well as those between two dsDNAs, in order to directly establish the relationship between the ion-mediated effective force and the ion-binding structures.

Several polyelectrolyte theories have been developed and employed to predict the ion-mediated interaction between like-charged polyelectrolytes. The counterion condensation theory has been developed based on line-charge structural model of infinite length, while the theory always predicts effective attractive forces between dsDNAs even at low monovalent salt and is only applicable to line-charge polyelectrolytes (33,34). The Poisson-Boltzmann theory with mean-field assumption ignores the ion-ion correlations and thus always predicts effective repulsive force between dsDNAs even at high multivalent salt (35-41). The electrostatic zipper model can predict an effective force between two helices at multivalent salt while the partition of binding ions into major/minor grooves is somewhat *ad hoc* (42,43). The tightly bound ion model which accounts for ion-ion correlation can predict the ion-mediated effective force between dsDNAs, while the model assumes the distribution of molecular charges on phosphate groups and thus may not make reliable predictions on ion-binding structures near helix grooves (44-47). Therefore, the existing theories could not give an *ab initio* and direct illustration on the multivalent ion-mediated effective interaction between dsRNAs (and dsDNAs) and the microscopic ion-binding structure. As an important bridge between experiments and theories, the computer simulations can be a powerful tool to probe the effective interactions between biomolecules (48-57). Beyond the simplified Monte Carlo and Langevin dynamics simulations (e.g., 47-50), the all-atomistic molecular dynamics (MD) explicitly takes into account the detailed atomistic structure of nucleic acids, ions, and water molecules, and thus could give a direct and detailed exploration on the effective interactions between biomolecules (48-63).

In this study, we will employ the all-atomistic MD simulations to directly calculate the potentials of mean force (PMF) between A-form RNAs (A-RNAs) and those between B-form DNAs (B-DNAs) in Co-Hex salt solutions. We will also make further calculations for A-form DNAs (A-DNAs) and for the A-RNAs with the spatial ion-accessible region modified. Correspondingly, the detailed Co-Hex binding structures will be analyzed. We will also make the direct comparisons with the related experiments. The present work would give atomistic level calculations on the effective interactions between dsRNAs as well as those between dsDNAs, and would present a direct illustration on the relationship between ion-mediated effective interactions and ion-binding structures.

## 2. Model and method

**2.1 All-atomistic molecular dynamics simulations**

In the work, we will calculate the PMFs between two A-RNAs, between two B-DNAs and between two A-DNAs by the atomistic MD simulations. The A-RNA, B-DNA and A-DNA helices are of 16-bp length, and their atomic structures are displayed in Fig. 1. The sequences of the helices are selected according to the recent experiments (30-32) and contain all the dinucleotide base pairs; see Table S1 in Supplementary Material.

The structures of the 16-bp nucleic acid helices are taken as the standard A-RNA, B-DNA and A-DNA fibers. The two parallel A-RNA (or DNA) helices with axes in *z* direction are separated in $x$ direction and immersed in a rectangle box containing explicit water and ions. The A-RNAs (or DNAs) are harmonically restrained with 1000 kJ/(mol.nm$^2$) force constant in $y$ and $z$ directions, thus are only allowed to move translationally in $x$ direction. The size of the rectangle box is taken as 130Å×80Å×100 Å. The Na$^+$ and Co(NH$_3$)$_6^{3+}$ (Co-Hex) ions are added with Amber tleap Program (64,65). To get the desirable bulk ion concentrations, the simplified Monte Carlo simulations (50,66) are employed to estimate numbers of Co-Hex and Na$^+$ ions in the simulational cell before the all-atomistic MD simulations; see Supplementary Material for details. Afterwards, the numbers of Co-Hex and Na$^+$ ions from the simplified Monte Carlo simulations are used in the all-atomistic simulations, and the realistic bulk Na$^+$ and Co-Hex concentrations from the all-atomistic MD simulations are close to the desirable values; see Fig. S1 in Supplementary Material for the cases of 100mM Na$^+$ and 5mM Co-Hex solutions.

In the simulations, we employ the Amber parmbsc0 force field and the TIP3P water model combined with parmbsc0 ion model (64,65). Corresponding to the recent experiments (30-32), the solutions always contain ~100mM NaCl as the background, and [Co-Hex] is taken as low (0.5mM), high (5mM) and very high (50mM) values, respectively. Since previous experiments showed that the octahedral coordination shell remains intact in binding to DNA (67), the Co-Hex ions are built with the explicit bonds between cobalt and amine groups and the explicit N–Co–N angles specified to generate an octahedral Co-Hex complex (68,69). The charges on a Co-Hex complex are generated based on the electrostatic potential generated with the DGAUSS program (68), and the hydrogen bonds of Co-Hex are considered with van der Waals potential with the new charge model and improved van der Waals parameters (68,70). All the parameters for Co-Hex ions are taken from Ref. (68). All the systems are optimized, thermalized (298K) and equilibrated by the program Gromacs 4.5 (71) with the periodic boundary conditions and Particle Mesh Ewald method employed (72), and a time step of 1-2fs is used in the conjunction with LINCS algorithm (73). Firstly, an energy minimization of 5,000 steps is performed with the steepest descent algorithm at low temperature, and then the systems are slowly heated to 298K and equilibrated with the Nose-Hoover temperature coupling until 0.5 ns (74). Afterwards, the subsequent NPT simulations of 2 ns (time-step 1fs, P=1atm) are performed with the Parrinello-Rahman pressure coupling and with the nucleic acids fixed (65). Finally, each MD simulation is continued for another 60ns in the isothermic-isobaric ensemble (time-step 2fs, P=1atm, T=298K). Our MD simulations generally reach the equilibrium after ~10ns, as shown in Figs. S2 and S3 for ion-binding number and Figs. S4 and S5 for helix-helix separation versus MD time in Supplementary Material. The trajectories in equilibrium are used to calculate the PMF between two helices.

## 2.2 Calculating potential of mean force between two helices

In the work, we employ the pseudo-spring method (53,54,66) to calculate the PMF between two A-RNAs (or DNAs). In the method, a pseudo-spring with spring constant $k$ is added to link the center of mass of two helices in MD simulations, as shown in Fig. 1. Based on the MD trajectories in equilibrium, the effective force between the two A-RNA (or DNA) helices can be calculated by

$$F = k\Delta x, \qquad (1)$$

where $\Delta x$ is the deviation of the spring length away from the original length $x_0$ in equilibrium. The negative and positive $\Delta x$'s correspond to the attractive and repulsive forces, respectively. After a series of $F(x)$ at different separations $x$ are obtained, the PMF $\Delta G(x)$ between the two A-RNA (or DNA) helices can be calculated through the integration (53,54,66)

$$\Delta G(x) = G(x) - G(x_{\text{ref}}) = \int_{x}^{x_{\text{ref}}} F(x') dx', \qquad (2)$$

where $x_{\text{ref}}$ is the outer reference separation. It has been shown previously that the pseudo-spring method is efficient and convenient in calculating PMF between two DNAs and two like-charged nanoparticles (53,54,66). In practice, the spring constant $k$ is generally taken as 1000 kJ/(mol.nm$^2$) and $x_{\text{ref}}$ is taken as 40Å. For the cases that two helices interact very strongly, we also use a higher value of $k$=2000 kJ/(mol·nm$^2$). Additionally, we have also made the additional calculations for the PMFs with the umbrella sampling method (70,71,75), and the results are very close to those from the pseudo-spring method; see Fig. S6 in Supplementary Material.

## 3. Results and discussion

In the work, we will calculate the PMF $\Delta G(x)$ between two nucleic acid helices in Co-Hex solutions by the all-atomistic MD simulations, and will examine how Co-Hex ions modulate the PMFs between dsDNAs and dsRNAs. We would emphasize illustrating the microscopic mechanism for the difference in PMFs between A-RNAs and B-DNAs.

### 3.1 PMFs between B-DNAs at low and high [Co-Hex]s

As shown in Fig. 2a, the PMF between two B-DNAs is repulsive at low (~0.5mM) [Co-Hex], while becomes strongly attractive with the free energy minimum of ~-3.5$k_B$T at the axis-axis separation of ~27Å when [Co-Hex] is increased to ~5mM. Such ion-mediated PMF is generally coupled to ion-binding (45,76). Due to the high entropy penalty for Co-Hex binding at low [Co-Hex], the system at low [Co-Hex] is dominated by Na$^+$-binding from the background of 100mM Na$^+$ and the monovalent ions can only modulate repulsive PMF. In contrast, at high [Co-Hex], Co-Hex-binding would dominate the system due to lower entropy ion-binding penalty, and strong Coulomb attraction between Co-Hex and two adjacent B-DNAs could cause an attractive force. The predicted PMFs between B-DNAs are in accordance with the previous experiments which show that high [Co-Hex] could induce DNA aggregation while DNAs resist condensation at low [Co-Hex] (12-17,30,31). The axis-axis separation of ~27Å at the lowest $\Delta G$(x) for ~5mM [Co-Hex] also agrees well with the values of equilibrium spacing of DNA aggregates from various experiments (15-19,77). In addition, following Refs. (46,78), we have calculated the osmotic pressures for hexagonal DNA aggregate, with assuming the additivity for the pair-wise PMF (28). The calculated osmotic pressures are very close to the corresponding experimental data (15,16); see Fig. S8 and Supplementary Material for details.

To gain a deep understanding on the relationship between [Co-Hex] and the PMF between B-DNAs, we would analyze the radial Co-Hex concentration distributions $c(r)$'s around B-DNAs at different [Co-Hex]'s, and $c(r)$'s have been calculated according to Eq. 9 in Ref (79). As shown in Fig. 2b, Co-Hex ions would begin to bind to a B-DNA at axial distance>~5Å, and prefer to accumulate at the outer surface of B-DNA with the axial distance of ~13Å. Such binding near the outer surface of a helix with the radial distance range of [11-15Å] will be termed as "external binding" (31). Higher (~5mM) [Co-Hex] causes much more Co-Hex ions "external binding" around B-DNAs near the radial distance of ~13Å than low (~0.5mM) [Co-Hex]. The further detailed analyses show that at low [Co-Hex], Co-Hex ions prefer to bind over the minor groove rather than the major groove, while at high [Co-Hex], Co-Hex ions mainly bind over the major groove rather than the minor groove; see Fig. 2c. This is reasonable since the minor groove of B-DNA is narrower and electrically more negative, thus Co-Hex ions prefer to binding to minor groove at low [Co-Hex] (31). When [Co-Hex] is increased to a high value, more Co-Hex ions become binding while the narrow minor groove has already accommodated many binding ions, causing the binding of excess Co-Hex ions over the major groove. The strong "external binding" of Co-Hex would be shared by adjacent B-DNAs, and could cause an significantly attractive PMF between B-DNAs at high [Co-Hex] (46,53,55,66,80).

To analyze the 3-dimensional Co-Hex binding around two B-DNAs, we would illustrate the binding structure of Co-Hex ions of high concentration (74). As shown in Figs. 3ab, Co-Hex ions form the obvious ion-bridge configuration between adjacent B-DNAs which appears much more pronounced for high [Co-Hex]. Such apparent ion-bridge configuration between two helices would provide a key contribution to the attractive PMF between B-DNAs (46,50,53,55,66).

At atomistic level, we have analyzed the structure of water molecules around bridging Co-Hex between two DNAs when two DNAs strongly attract each other. As shown in Fig. S9 in Supplementary Material, bridging Co-Hex can induce the ordering of water molecules between adjacent phosphates in two DNAs, i.e., Co-Hex can induce the rotation of $H_2O$ with O atoms pointing to Co-Hex and H atoms pointing to phosphates. Such configuration of bridging Co-Hex-induced ordering of water molecules between two helices could mediate an apparent DNA-DNA attraction, as suggested by Parsegian, Rau, Qiu and coworkers (13,19).

**3.2 PMFs between A-RNAs at low, high and very high [Co-Hex]s**

The PMFs between A-RNAs have been calculated at low (~0.5mM), high (~5mM) and very high (~50mM) [Co-Hex]'s, as shown in Fig. 2d. At low (~0.5mM) [Co-Hex], in analogy to B-DNAs, the PMF between A-RNAs is repulsive and appears slightly stronger than that between B-DNAs. This may be attributed to the slightly higher charge density and thicker helix of A-RNAs (1). As [Co-Hex] is increased to ~5mM, the PMF between A-RNAs is still (weakly) repulsive with weaker strength than that at low [Co-Hex], which is distinctively in contrast to the strongly attractive PMF between B-DNAs. With the increase of [Co-Hex] to a very high value (~50mM), the PMF between A-RNAs becomes strongly attractive with free energy minimum of ~-4.1$k_BT$ at axis-axis separation of ~27Å. The predicted PMFs at different [Co-Hex]'s are in good accordance with the recent experiments on dsRNAs which have shown that dsRNAs resist condensation when [Co-Hex] <~5mM while could condense at very high [Co-Hex], relatively to the background $Na^+$ (30,31).

To understand the [Co-Hex]-dependent PMF between A-RNAs, we have analyzed the radial Co-Hex concentration distribution around A-RNAs (79), as shown in Figs. 2ef. For convenience, corresponding to the above termed "external binding" with the radial distance range of [11-15Å] (31), another binding mode with the radial distance range of <~11Å (~helical radii of dsDNA and dsRNA) is termed as "internal binding" (31). Overall, Co-Hex ions begin to bind to A-RNA at radial distance >~2 Å, which is attributed to the accessible deep major groove of A-RNA. This is distinctly different from B-DNA. At low (~0.5mM) [Co-Hex], Co-Hex ions exhibit "internal binding" into the deep major groove within the radial distance of <~10Å. As [Co-Hex] is increased to ~5mM, Co-Hex ions prefer to bind internally into the deep major groove around the radial distance of ~8Å, and Co-Hex binding distribution is extended to the radial distance range of <~12Å. With the increase of [Co-Hex] to ~50mM, Co-Hex ions would still show preference to bind internally into the major groove around radial distance of 8-9Å, while the apparent binding distribution is extended to the radial distance range of <~16Å and the [Co-Hex] at ~15Å can nearly reach ~0.3M. The concentration-dependent Co-Hex-binding distribution around A-RNA is understandable. At low [Co-Hex], Co-Hex would prefer to bind into the deep major groove with small radial distance where electric field is strongest (31), and as [Co-Hex] is increased, more Co-Hex ions become binding and begin to externally bind near outer surface of A-RNA after the deep/narrow major groove is fulfilled. Very high (e.g., ~50mM) [Co-Hex] would cause apparent "external binding", and such apparent "external binding" of Co-Hex can be shared by adjacent A-RNAs to cause an attractive PMF at very high [Co-Hex].

The 3-dimensional Co-Hex binding around two A-RNAs has been directly illustrated in Figs. 3e-f. At low [Co-Hex], Co-Hex ions of high density are "internal binding" in the deep and narrow major groove, and there is no visible ion-bridge between two A-RNA surfaces. At high (~5mM) [Co-Hex], the abundant Co-Hex ions of high density reside in the major groove, and there is still no evident ion-bridge, which corresponds to the (weakly) repulsive PMF between A-RNAs. When [Co-Hex] is increased to ~50mM, many more Co-Hex ions become binding and the excess Co-Hex would bind externally at outer surface of A-RNAs, and the apparent ion-bridge between two A-RNAs is formed. Such ion-bridging configuration with Co-Hex-induced water ordering would be responsible for the strongly attractive PMF between A-RNAs at very high [Co-Hex] (46,50,53,55,66).

### 3.3. PMF is dependent on helical structure: dsRNA versus dsDNA

*3.3.1. A-RNA versus B-DNA*

Both A-RNAs and B-DNAs are highly (negatively) charged polymers, while the effective interaction between A-RNAs appears distinctly different from that between B-DNAs at high (~5mM) [Co-Hex], i.e., the PMF between B-DNAs is strongly attractive while that between A-RNAs is repulsive; see Fig. 4a and Figs. 2ad. As shown above, there are distinct differences between A-RNA and B-DNA in their topological structures. A-RNA has the deep/narrow major groove and the wide/shallow minor groove, as shown in Fig. 3. Since there is strongest electric potential in the deep/narrow major groove of A-RNA, Co-Hex would prefer to bind deeply into major groove. Such "internal binding" cannot contribute to the formation of ion-bridge between two A-RNAs and consequently the PMF is repulsive. Unlike A-RNA structure, B-DNA has the wide major groove and the narrow minor groove which are not deep compared with those of A-RNA, and Co-Hex ions would like to become "external binding" near outer surface of B-DNA rather than "internal binding" deeply into grooves. Such "external binding" would help to form the apparent ion-bridge which could be shared by adjacent B-DNAs to result in an attractive PMF. When [Co-Hex] is increased to a very high value, the excess Co-Hex can also "externally bind" to A-RNA since the deep/narrow major groove is fulfilled by more binding Co-Hex ions. Such "external binding" induced by a very high [Co-Hex] can also promote the formation of ion-bridge and cause a strong attractive force between A-RNAs.

*3.3.2. PMF between A-DNAs*

To further confirm the above described effects of "internal binding" and "external binding", we have calculated and analyzed the PMF between two A-DNAs at 5mM [Co-Hex]. As shown in Fig. 4a, in analogy to A-RNA, the PMF between A-DNAs is repulsive at 5mM [Co-Hex]. The detailed analysis on Co-Hex binding distribution also shows the "internal binding" near the radial distance of ~8Å, which is similar to that of A-RNA; see Figs. 4bc. As illustrated in Fig. 3c, such "internal binding" is in deep major groove and resists the formation of ion-bridge, which is responsible for the (weakly) repulsive PMF between A-DNAs. It is noted here that despite the similarity in "internal binding", the radial distribution of Co-Hex around A-DNA is slightly different from that around A-RNA: Co-Hex concentration around A-RNA is higher at radial distance of <~5 Å, while lower at radial distance of ~8 Å than that around A-DNA; see Figs. 4bc. Such difference comes from the difference in helical structures of A-RNA and A-DNA. Despite the similar A-form helical structure, as indicated in the experiments (81-83), the width of major groove of A-DNA is ~2Å narrower than that of A-RNA, which is also in accordance with our used structure parameters. The narrower major groove of A-DNA would cause Co-Hex ions to have more preference to bind in major groove and coordinate with adjacent phosphate strands at the radial distance of ~8Å, which has been observed in our MD simulations and illustrated in Fig. 3c. The radial profiles of Co-Hex concentrations around A-form helices are in accordance with the previous experiments on Co-Hex binding to A-DNA, which showed that Co-Hex would bind to bases in major groove and to phosphates, either bridging across narrow major groove or residing between two adjacent intra-strand phosphates (84).

*3.3.3. PMF between "modified" A-RNAs*

Since it is the "internal binding" that resists the formation of ion-bridge and causes the repulsive PMF between A-RNAs, there would be an interesting question: If Co-Hex ions are prohibited to enter deeply into the major groove of A-RNAs, can A-RNAs attract each other? To answer the interesting question and further validate the above analysis, we have made the further calculations for the "modified" A-RNAs (A-RNA' and A-RNA") at 5mM [Co-Hex]. A-RNA' corresponds to the A-RNA with the bottom of the central 1/3 major groove fixed by a layer of water, and A-RNA" corresponds to the A-RNA with the bottom of the whole major groove fixed by a layer of water, as shown in Fig. 5. Such treatment of fixing water at the bottom of deep major groove may effectively exclude the very deep binding of Co-Hex in major groove, and may possibly cause the "external binding" and strong attractive force between the modified A-RNAs. Our calculations show that there are indeed apparently attractive forces between the "modified" A-RNAs; see Fig. 6a. The analyses on Co-Hex concentration distributions around the "modified" A-RNAs also show that the treatment of excluding Co-Hex from the deep major groove promotes the "external binding". As shown in Figs. 6bc, when a layer of water is fixed at the bottom of major groove, the original "internal binding" in the radial distance range of ~2.5Å to ~12.5 Å would change into the binding in the radial distance range of ~7Å to ~12.5 Å, and the "external binding" near the radial distance of ~13Å. Furthermore, the direct illustration in Fig. 5 shows that such treatment would definitely help to form the ion-bridge between two A-RNAs, which could cause the strong attractive PMF between the modified A-RNAs (46,50,53,55,66).

## 4. Conclusion

In this work, the atomistic MD simulations have been employed to calculate and analyze the potentials of mean force between A-RNAs, between B-DNAs, and between A-DNAs in Co-Hex solutions. The present work shows that though the nucleic acid helices have similar negative charge densities, the effective interactions between them are distinctively different. At high [Co-Hex], two B-DNAs strongly attract each other, while two A-RNAs and two A-DNAs repel each other. The present analysis shows that such significant difference between B-form and A-form helices is attributed to the ion-binding structure. For B-DNA, Co-Hex ions would become "external binding" around phosphates and form the ion-bridge between two B-DNAs. But for A-RNA and A-DNA, due to the existence of deep major groove, Co-Hex would preferentially exhibit "internal binding" into the major groove and consequently cannot form the ion-bridge between two A-form helices, causing the repulsive interaction between them. The effective interactions between A-RNAs can become strongly attractive when Co-Hex ions become "external binding" to form the apparent ion-bridge, which can be realized by either increasing Co-Hex to a very high concentration or fixing a layer of water at the bottom of the deep major groove.

Overall, our results are in accordance with the experimental findings (17,30,31,77). First, our calculations show that the PMF between A-RNAs is weakly repulsive while that between B-DNAs is strongly attractive at ~5mM Co-Hex, which is in good agreement with the recent experiments (30,31). Second, Parsegian, Rau and coworkers combined single-molecule magnetic tweezers with osmotic stress on DNA assembly in various salt solutions. Their direct measurement of the free energy for DNA aggregates in Co-Hex solution is ~-0.21 $k_BT$/bp (17), a value close to ours of ~-3.5$k_B$T for 16-bp DNAs, and the separation of ~27Å at the lowest free energy for ~5mM Co-Hex from our MD simulations is also close to the experimental equilibrium spacing ~28 Å of DNA aggregate (15,17). Third, the calculated osmotic pressures with assuming additivity of PMF are very close to the corresponding experimental measurements (15,16).

Kornyshev and Leikin have successfully developed their electrostatic zipper model to explain why A-RNAs resist condensation while B-DNAs would become condensed, and such distinctive behaviors were proposed to be attributed to the different widths of the major grooves of different helices (42). However, the model has also involved some important simplifications such as assuming the uniform distribution of ions in grooves and ignoring ion size and 3-dimensional ion-accessible geometry of different helices. The present all-atom MD simulations with explicit helical structures, ions and water molecules, show that the 3-dimensional topology of different helices can play an important role in modulating PMF between two helices. Specifically, the attractive PMF between B-DNA at ~5mM Co-Hex is accompanied with the external binding of Co-Hex above major groove at radial distance of ~13Å, while the non-attractive PMF between A-RNAs at ~5mM Co-Hex is accompanied with the internal binding of Co-Hex at radial distance of ~8-9Å in major groove.

The present work has also involved some assumptions and simplifications. First, all the dsRNAs and dsDNAs have been approximately treated as rigid bodies and thus the stability and flexibility of ds helices have been ignored. Since all the nucleic acid helices are in solutions of 100mM NaCl, the helices would have high stability to keep its helical structure rather than become denatured (85,86). Although nucleic acid helices are flexible in ionic solutions, the ignorance of helix flexibility can be a reasonable approximation due to the high persistence length (45nm-60nm) (81,82) and strong stretching modulus (>~500pN) (74,87,88) of dsDNA and dsRNA. Second, we have made the calculations for A-DNAs without considering the likely dependence of A-DNA structure on ionic condition (1). It is a reasonable simplified treatment since we only use A-DNA as a structure model to further examine the mechanism. Thirdly, the extensive relevant experiments have used the 25-bp dsRNA and dsDNA helices (30,31), while 16-bp dsRNAs and dsDNAs have been employed in the work for saving the computational time. Such simplification would not affect the conclusions since the experiments have shown that 16-bp dsDNAs have the similar Co-Hex-dependent condensation behavior (30). Fourthly, the related experiments involve different background $Na^+$ concentrations of 100mM and 20mM with different experiment techniques (30,31), while the present work only considers the solutions containing 100mM $Na^+$. Such simplification should not qualitatively affect our conclusions since the experiments with different background $Na^+$ concentrations have shown the qualitatively similar results (30,31). The effect of competition between Co-Hex and background $Na^+$ on the effective nucleic acid helix-helix interaction may deserve to be studied separately.

Furthermore, in the present work, we only consider the parallel configuration of two helices with all the atoms restrained in y and z directions, and the helices cannot rotate around any of the x, y and z axes. In fact, such parallel configuration would be the most favorable one for large separation between two helices (89), while may become unfavorable for the closely packaging of two helices, due to the strong electrostatic repulsion (28,29). In realistic 3D space, the axes of two helices at small separation can rotate by a small angle to form a typical X-shaped structure (90,91) with several possible packaging modes for different helical structures (90,91). Such tight non-parallel helix-helix configuration may be important for the packaging of some RNAs such as the P4-P6 domain of the Tetrahymena thermophila intron (92).

Finally, the present work only involves the system of two helices while the related experiments generally involved multiple helices. The previous studies have shown that the multi-body effect may slightly affect the multivalent-mediated helix-helix attraction and consequently may slightly affect the comparisons between our predictions and experiments on the interaction strength and axis-axis separation (28). The strict and extensive exploration for multi-body PMF between helices at atomistic level is deserved to be studied separately.

Nevertheless, the present work has provided the direct calculations on the PMFs between A-RNAs, between B-DNAs and between A-DNAs, and has directly illustrated the microscopic mechanisms for the different PMFs between double-stranded nucleic acids with different helical structures.

## 5. Acknowledgements

We are grateful to Profs. Shi-Jie Chen, Haiping Fang and Wenbing Zhang for valuable discussions, and Chang Shu for facility assistance. This work was supported by the National Key Scientific Program (973)-Nanoscience and Nanotechnology (No. 2011CB933600), the National Science Foundation of China grants (11074191, 11175132 and 11374234), and the Program for New Century Excellent Talents (Grant No. NCET 08-0408). One of us (Y.Y. Wu) also thanks the financial support from the interdisciplinary and postgraduate programs under the Fundamental Research Funds for the Central Universities.

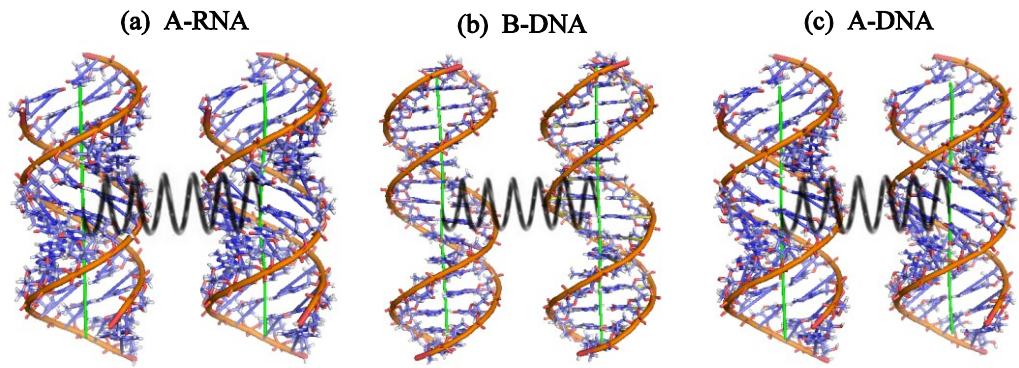

**Figure 1** An illustration for two parallel 16-bp A-RNAs (a), B-DNAs (b) and A-DNAs (c). The spring with a spring constant *k* which connects the centers of mass of two helices has been used to calculate the potential of mean force between two double-stranded RNAs and DNAs (53,54,73).

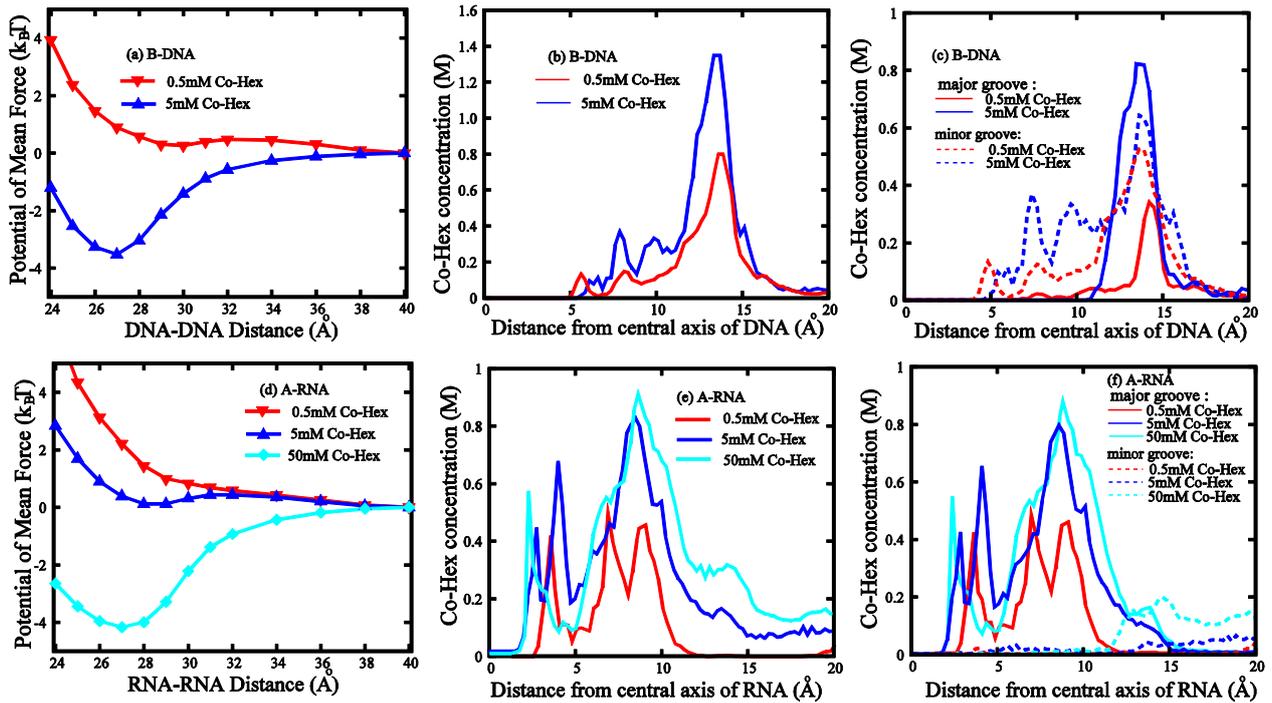

**Figure 2** (a) The potentials of mean force between two 16-bp B-DNAs in 0.5mM and 5mM Co-Hex solutions. (b) The Co-Hex ion distributions around the B-DNA corresponding to panel a. (c) The Co-Hex ion distributions around B-DNA in (or over) major groove and minor groove according to panels a and b. (d) The potentials of mean force between two 16-bp A-RNAs in 0.5mM, 5mM and 50mM Co-Hex solutions. (e) The Co-Hex distributions around the A-RNA corresponding to panel d. (c) The Co-Hex distributions around A-RNA in (or over) major groove and minor groove corresponding to panels d and e. Note that the buffers always contain 100mM NaCl.

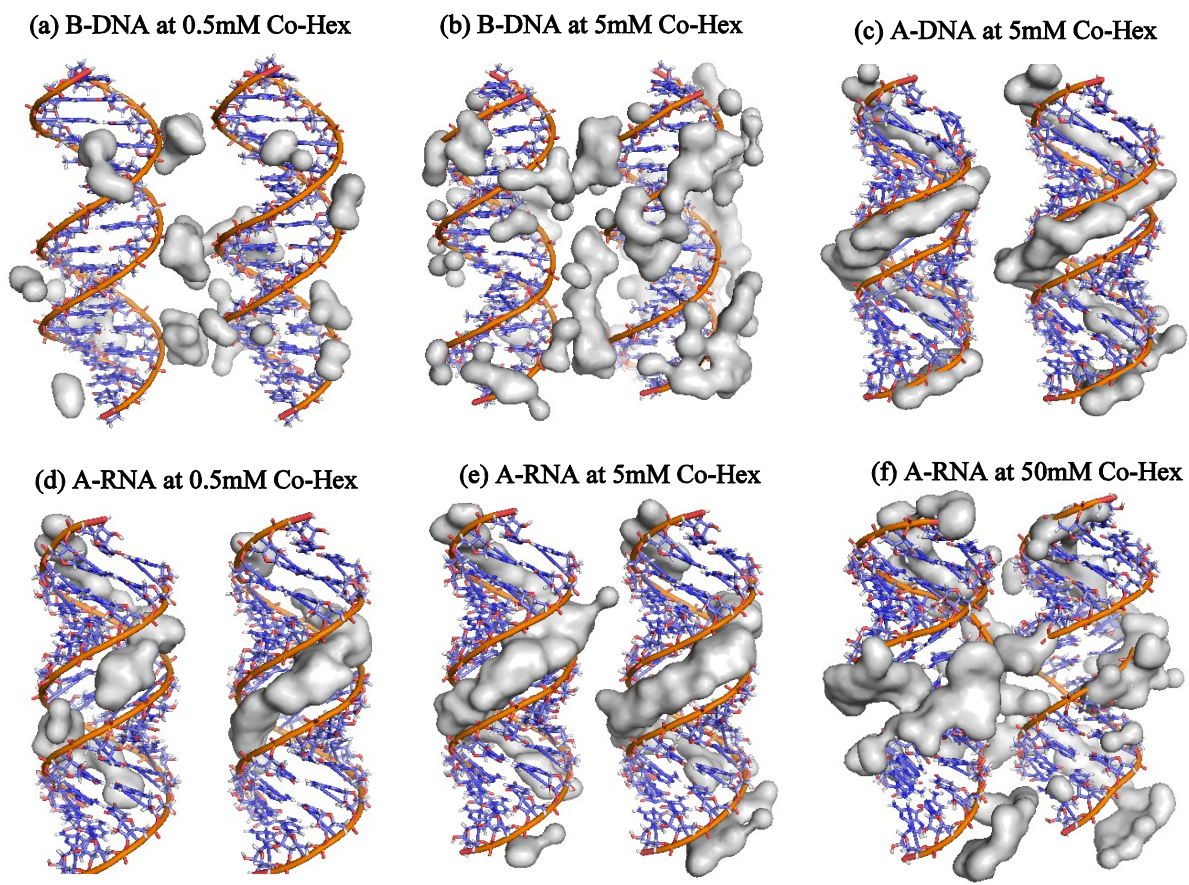

**Figure 3** (a,b) The illustrations for the region of high Co-Hex ion charge density (larger than 0.02 $e$/Å$^3$) around two 16-bp B-DNAs in 0.5mM (a) and 5mM (b) Co-Hex solutions; (c) The illustration for the region of the high Co-Hex ion charge density (larger than 0.02 $e$/Å$^3$) around two 16-bp A-DNAs in 5mM Co-Hex solutions. (d-f) The illustrations for the region of high Co-Hex charge density (larger than 0.02 $e$/Å$^3$) around two 16-bp A-RNAs in 0.5mM (d), 5mM (e) and 50mM (f) Co-Hex solutions. Note that the buffers always contain 100mM NaCl.

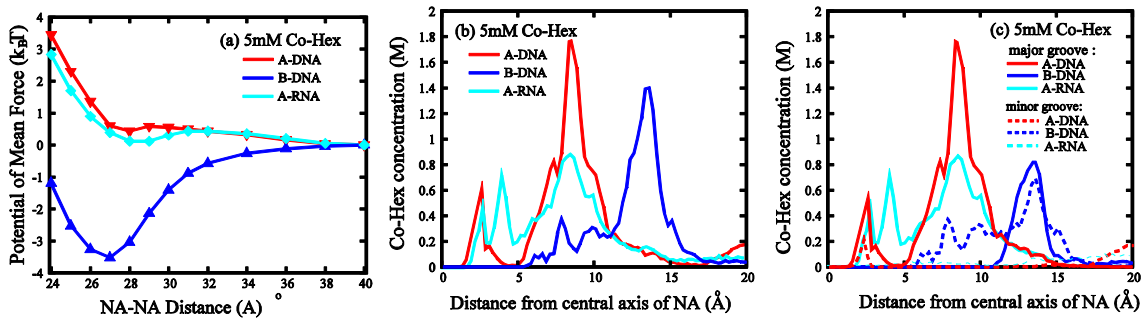

**Figure 4** (a) The potentials of mean force between two 16-bp nucleic acid (B-DNA, A-RNA, and A-DNA) helices in 5mM Co-Hex ion solution; (b) The Co-Hex distribution around the nucleic acid helices corresponding to panel a; (c) The Co-Hex distribution around the nucleic acid helices in (or over) major groove and minor groove corresponding to panels a and b. Note that the buffers always contain 100mM NaCl.

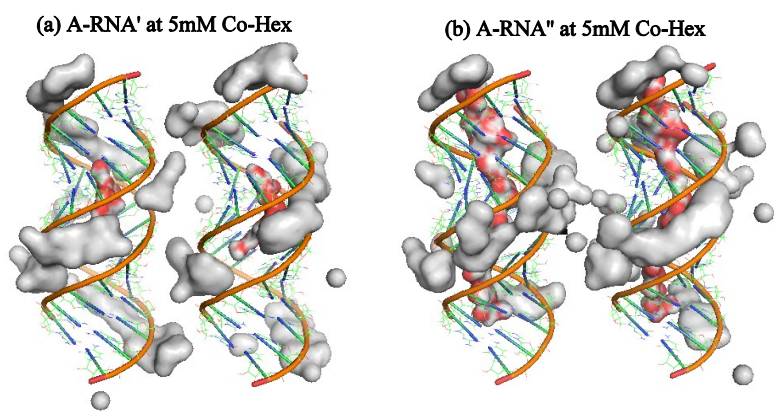

**Figure 5** The illustrations for the spatially modified A-RNA (A-RNA' and A-RNA") and the region of the high Co-Hex charge density (larger than 0.02 $e/Å^3$) in 5mM Co-Hex solution around two 16-bp A-RNA's and around two 16-bp A-RNA's. The A-RNA' denotes the A-RNA with the bottom of the central 1/3 major groove fixed with a layer of water, whereas the A-RNA" denotes the A-RNA with the bottom of the entire major groove fixed with a layer of water. The red-gray chains in deep major grooves illustrate the fixed water molecules. Note that the buffers always contain 100mM NaCl.

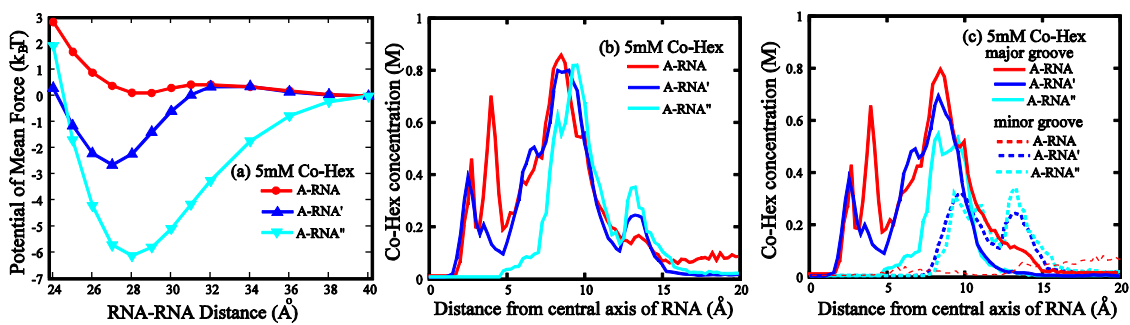

**Figure 6** (a) The potentials of mean force between two 16-bp A-RNAs, between two 16-bp A-RNA's, and between 16-bp A-RNA''s in 5mM Co-Hex ion solution; (b) The Co-Hex distributions around the A-RNA, A-RNA', and A-RNA'' corresponding to panels a and b; (c) The Co-Hex distributions around the A-RNA, A-RNA', and A-RNA'' in (or over) major groove and minor groove corresponding to panels a and b. The A-RNA' denotes the A-RNA with the bottom of the central 1/3 major groove fixed with a layer of water, whereas the A-RNA'' denotes the A-RNA with the bottom of the entire major groove fixed with a layer of water. Note that the buffers always contain 100mM NaCl.

Supplementary Material for

# Multivalent ion-mediated nucleic acid helix-helix interactions: RNA versus DNA

Yuan-Yan Wu[1], Zhong-Liang Zhang[1], Jin-Si Zhang[1], Xiao-Long Zhu[2] and Zhi-Jie Tan[1*]

[1]Department of Physics and Key Laboratory of Artificial Micro & Nano-structures of Ministry of Education, School of Physics and Technology, Wuhan University, Wuhan 430072, China

[2]Department of Physics, School of Physics & Information Engineering, Jianghan University, Wuhan 430056, China

*To whom correspondence should be addressed: zjtan@whu.edu.cn

**To get desirable bulk Co-Hex and Na$^+$ concentrations**

Due to the competition between monovalent and multivalent ions in binding to nucleic acids (e.g., Ref. (29)), it is not straightforward to obtain desirable bulk monovalent/multivalent ion concentrations in a MD simulation for a nucleic acid in a mixed monovalent/multivalent ion solution.

In the present work, to get the desirable bulk ion concentrations, before the all-atom MD simulations, the simplified Monte Carlo (MC) simulations (50,66) are employed to estimate numbers of Co-Hex and Na$^+$ ions in the simulational cell. Practically, all-atom structure of nucleic acids and ions are placed in the MC simulational cell, and water is treated as continuous medium with dielectric constant of 78 (50,66). The MC algorithm with Coulomb and Lenard-Jones potentials (50,66) is performed to get the bulk ion concentrations in equilibrium. We change the relative numbers of Co-Hex and Na$^+$ ions and repeat the MC processes, and then we can estimate the numbers of Co-Hex and Na$^+$ ions in the simulational cell at the desirable bulk ion concentrations. Afterwards, the numbers of Co-Hex and Na$^+$ ions from the simplified MC simulations are used in the all-atom simulations, and the realistic bulk Na$^+$ and Co-Hex concentrations from the all-atom MD simulations are very close to the desirable values; see Fig. S1 in Supplementary Material for the cases of 100mM Na$^+$/5mM Co-Hex solutions.

**Osmotic pressure for DNA aggregates and comparison with experimental data**

For DNA array (DNA aggregates), the osmotic pressures have been measured experimentally by the osmotic stress technique. In this section, based on the pair-wise DNA helix-helix interactions, we have calculated the osmotic pressures and compared the results with the experimental data. For a hexagonal DNA array (see Fig. 11a in Ref. (46)), the mean free energy Δg(x) per DNA can be approximately calculated through the summation over the pair-wise helix-helix interactions between nearest neighbor pairs (46,78)

$$\Delta g(x) = \sum_{i=1}^{6} \Delta G_i(x)/2, \quad (S1)$$

where $\Delta g(x) = \sum_{i=1}^{6} \Delta G(x)$ is the total free energy between an helix and its six neighbors, and the factor 1/2 is used to remove double-counting. $\Delta G_i(x)$ is the free energy for the two-helix system of a helix and its i-th neighbor. Here, we have neglected the nonadditive effect in multiple helix packing and kept only interactions between the nearest neighboring helices, i.e., when calculating the interaction between two helices, we ignore the existence of other helices.

The osmotic pressure Π(x) as a function of the helix-helix distance x can be calculated from (46,78)

$$\Pi(x) = -\frac{\Delta g(x)}{\partial V}, \quad (S2)$$

where V=L×A is the volume of the hexagonal region around each helix. L is the length of each helix, and $\sqrt{3}x^2/2$ A=3 is the average cross section area per molecule in the DNA array (46). Practically, we first fit the calculated PMF to a polynomial function and based on the fitted PMF, we could easily calculate the osmotic pressures according to Eqs. S1 and S2.

As shown in Fig. S8, for DNAs in 5mM Co-Hex/100mM $Na^+$, the calculated osmotic pressures are very close to the experimental data (15,16), and the slight deviation may come from the multi-helix effect (46) which was ignored in our calculations for hexagonal DNA aggregate. For DNAs in 0.5mM Co-Hex/100mM $Na^+$, there is no directly available experimental data. As shown in Fig. S8, the addition of 0.5mM Co-Hex in 100mM $Na^+$ can cause a different osmotic pressure curve from that for 100mM $Na^+$ (Ref. (1) in supplementary material).

**Table S1** A-RNA, B-DNA and A-DNA sequences used in the present study.[a]

| Nucleic Acids | Sequences |
|---|---|
| A-RNA | 5'-CGACUCUACUACGCGC-3' <br> GCUGAGAUGAUGCGCG |
| B-DNA | 5'-CGACTCTACTACGCGC-3' <br> GCTGAGATGATGCGCG |
| A-DNA | 5'-CGACTCTACTACGCGC-3' <br> GCTGAGATGATGCGCG |

[a]The sequences of the short RNA and DNAs are selected according to the recent small angle scattering experiments and UV measurements (29).

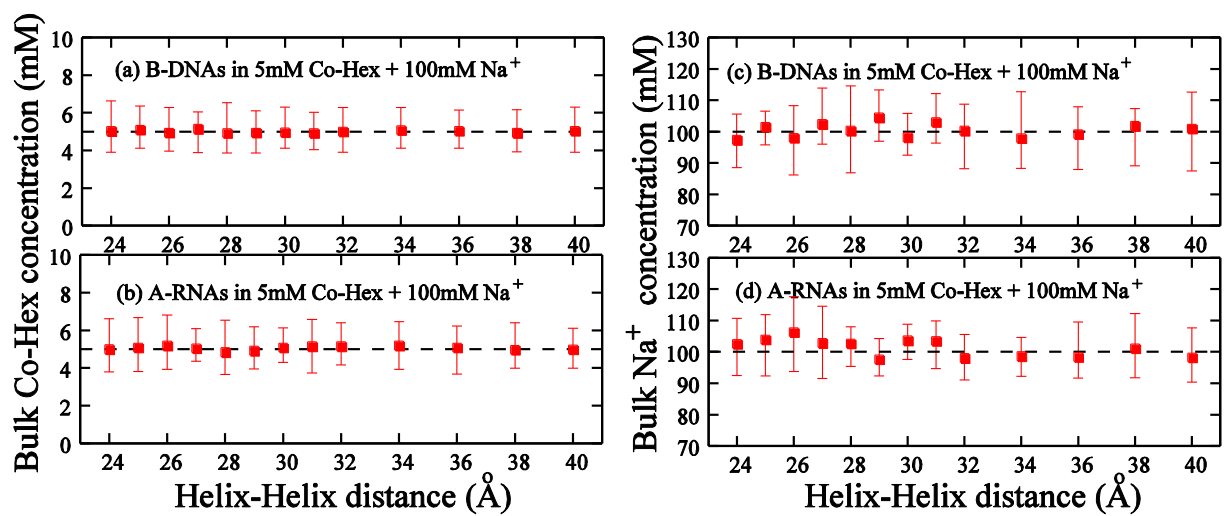

**Figure S1** (a,c) The realistic bulk Co-Hex (a) and $Na^+$ (c) concentrations for two DNAs in 100mM $Na^+$/5mM Co-Hex solutions from the all-atom MD simulations as a function of DNA-DNA distance; (b,d) The realistic bulk Co-Hex (b) and $Na^+$ (d) concentrations for two RNAs in 100mM $Na^+$/5mM Co-Hex solutions from the all-atom MD simulations as a function of RNA-RNA distance.

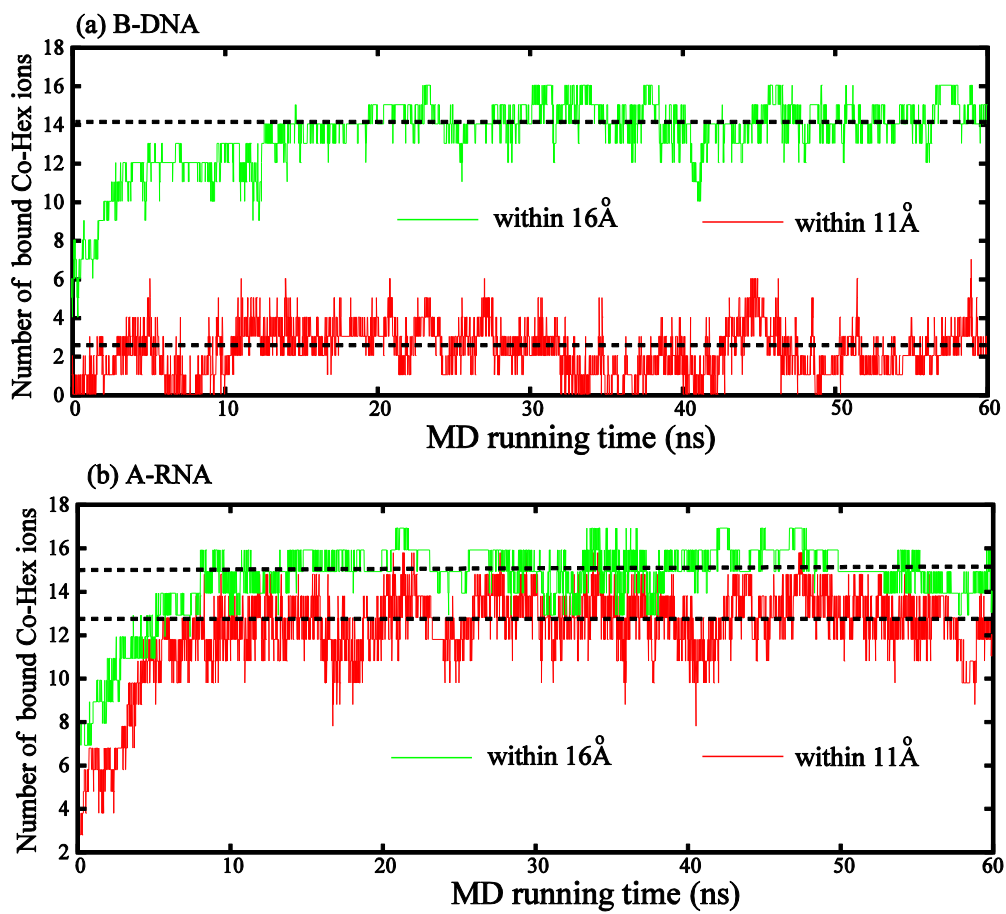

**Figure S2** Number of bound Co-Hex ions within the distances of 11Å (red) and 16Å (green) from two duplex helical axes as function of MD running time: (a) B-DNAs and (b) A-RNAs at 5mM Co-Hex and 100mM Na$^+$ ion solution. Dashed lines denote the averaged values in equilibrium.

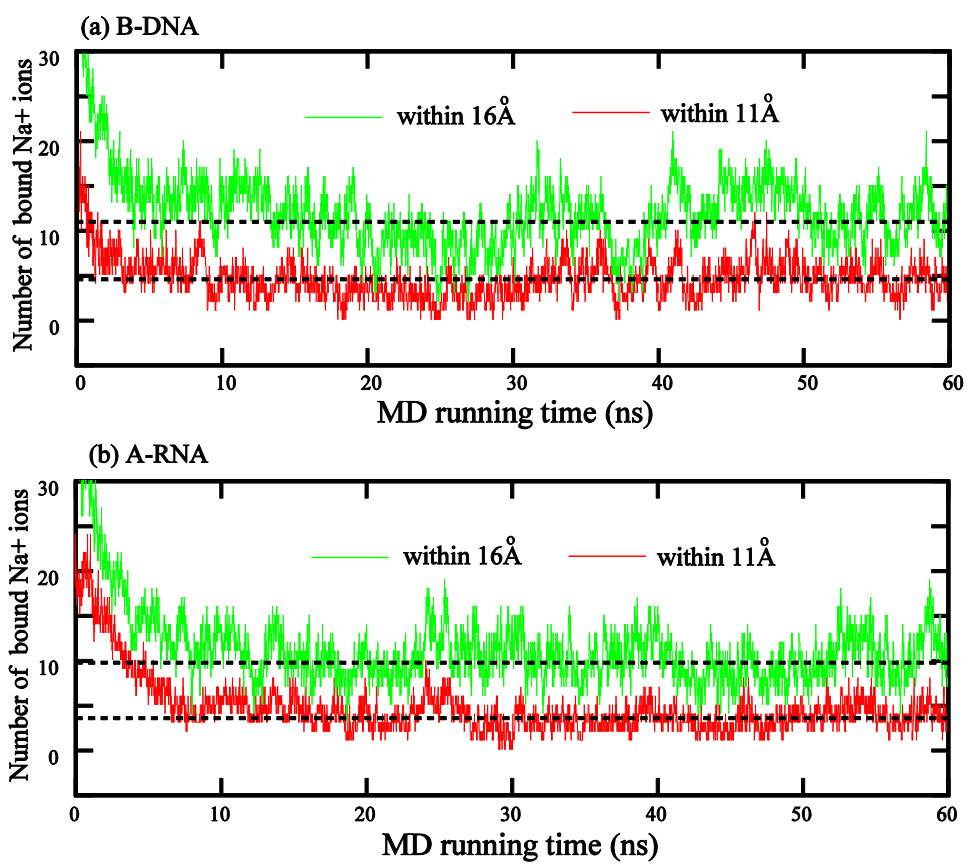

**Figure S3** Number of bound Na$^+$ ions within the distances of 11Å (red) and 16Å (green) from two duplex helical axes as function of MD running time: (a) B-DNA and (b) A-RNA at 5mM Co-Hex and 100mM Na$^+$ ion solution. Dashed lines denote the averaged values in equilibrium.

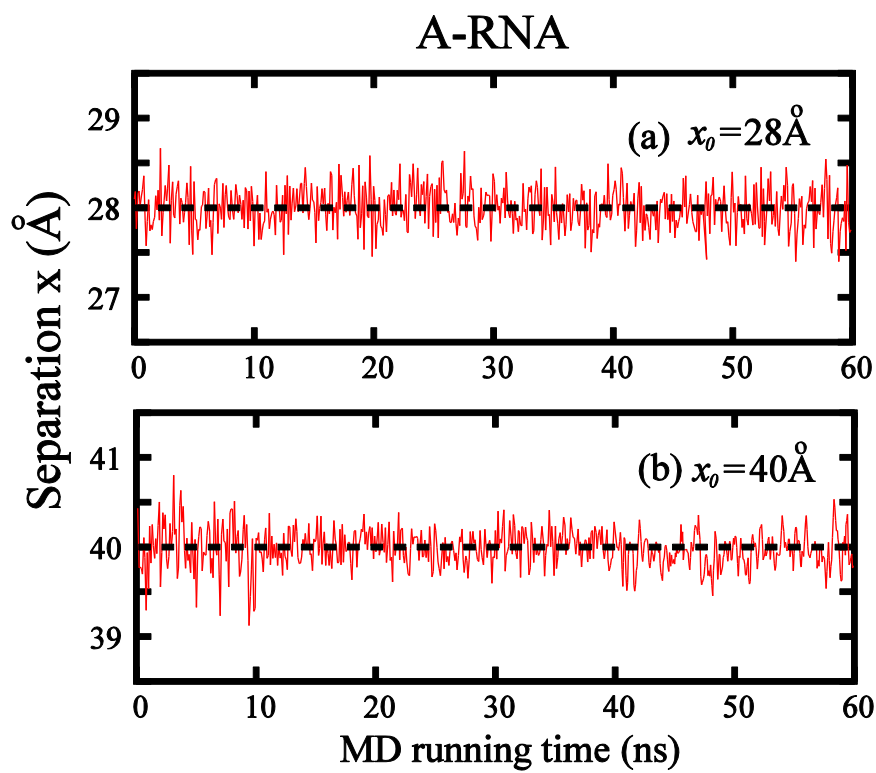

**Figure S4** An illustration of the MD convergence of separation $x$ between the centers of mass of two 16-bp A-RNA helices in 5mM Co-Hex ion solution with 100mM NaCl. The averaged values of separation $x$ are made over every $\Delta t$ by $\int_{t-\Delta t/2}^{t+\Delta t/2} x(t')dt'/\Delta t$ and $\Delta t =20$ps. The black lines denote the averaged values in equilibrium. Two typical original separations $x_0$ between two A-RNAs are shown in the panels.

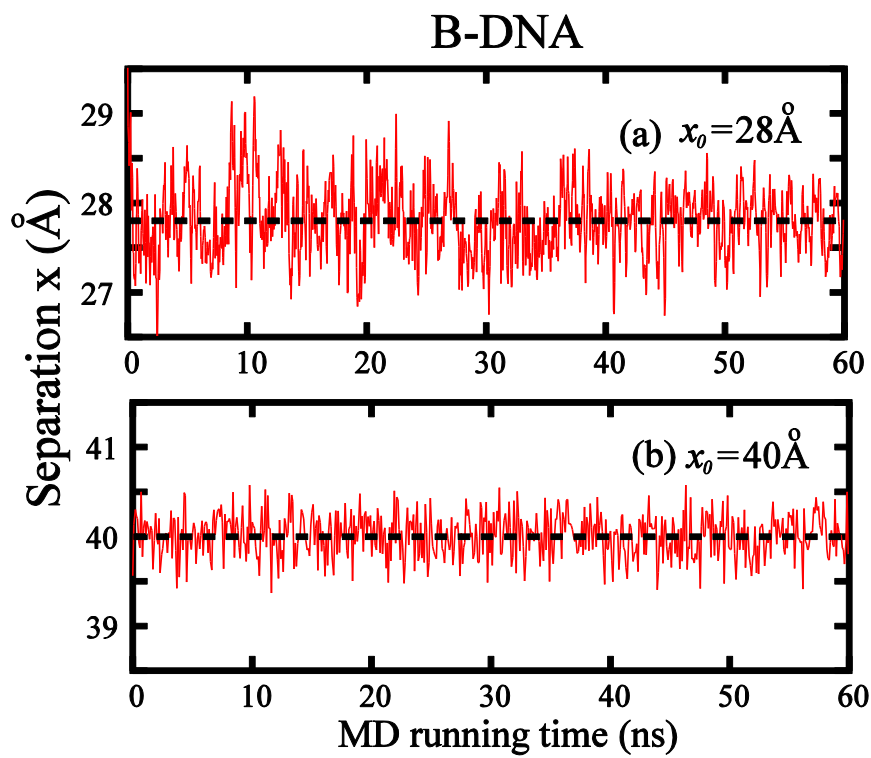

**Figure S5** An illustration of the MD convergence of separation $x$ between the centers of mass of two 16-bp B-DNA helices in 5mM Co-Hex ion solution with 100mM NaCl. The averaged values of separation $x$ are made over every $\Delta t$ by $\int_{t-\Delta t/2}^{t+\Delta t/2} x(t')dt'/\Delta t$ and $\Delta t$ =20ps. The black lines denote the averaged values in equilibrium. Two typical original separations $x_0$ between two A-DNAs are shown in the panels.

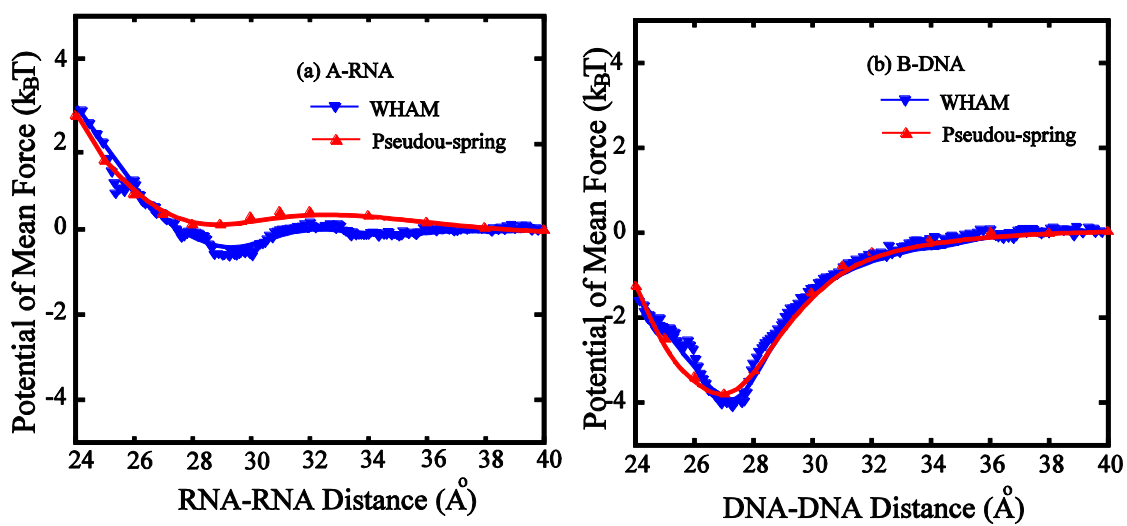

**Figure S6** The potential of mean force as a function of the separation *x* between the centers of mass of two 16-bp nucleic acid helices in 5mM Co-Hex ion solution with 100mM NaCl: (a) A-RNAs and (b) B-DNAs. Blue: calculated from the umbrella sampling with the weighted histogram analysis method (WHAM); Red: calculated from the pseudo-spring method employed in the present work.

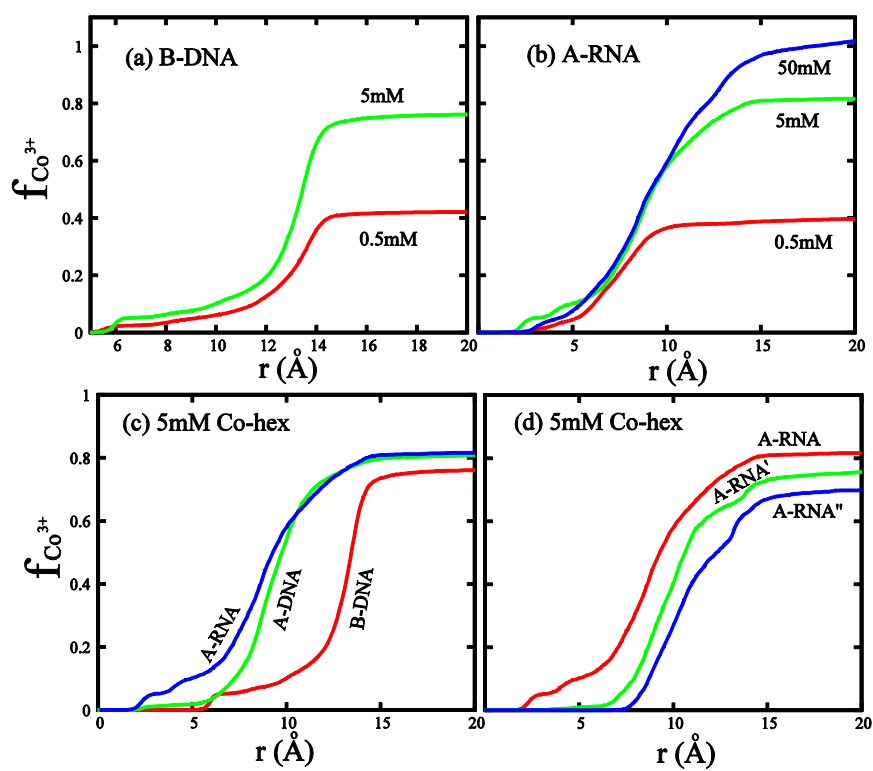

**Figure S7** The Co-Hex charge distribution per unit charge on nucleic acid helices as a function of distance *r* from the axes of nucleic acid helices. (a) B-DNAs at 0.5mM and 5mM [Co-Hex]'s; (b) A-RNAs at 0.5mM, 5mM and 50 mM [Co-Hex]'s; (c) The comparisons between A-RNAs, B-DNA sand A-DNAs; (d) The modified A-RNAs. The A-RNA' denotes the A-RNA with the bottom of the central 1/3 major groove fixed with a layer of water, whereas the A-RNA" denotes the A-RNA with the bottom of the entire major groove fixed with a layer of water. Please note that the buffers always contain 100mM NaCl.

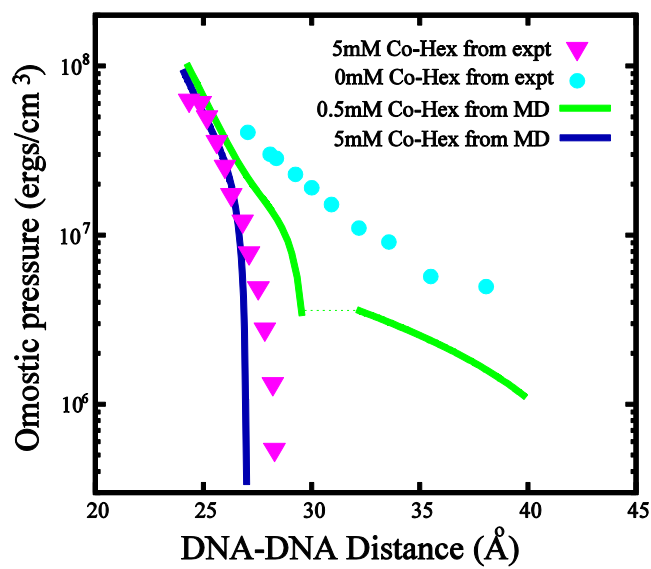

**Figure S8** Comparison of osmotic pressure curves for hexagonal DNA aggregate in various salt solutions from MD simulations (lines) and experiments (symbols). Lines: DNAs at 0.5mM (green) and 5mM (blue) [Co-Hex]'s, and the buffers always contain 100mM NaCl. Symbols: ▼, experimental data for DNAs at 5mM Co-Hex and 100mM NaCl with 10mM TrisCl (17); •, experimental data for DNAs at 100mM NaCl with 10mM TrisCl (Ref. 1 in Supplementary Material). The apparent deviation between blue curve and • comes from that the experiments did not involve Co-Hex ions.

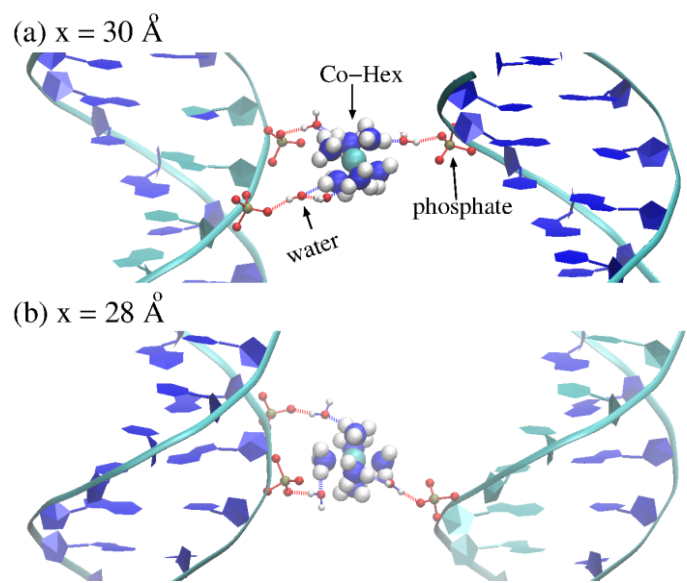

**Figure S9** Two snapshots to illustrate the structure of water molecules which link the bridging Co-Hex and phosphates in two DNAs. The axis-axis distances between DNAs are 30Å (a) and 28 Å (b), respectively. The dash lines are H-bonds which are automatically displayed by the software VMD (Ref. 2 in Supplementary Material).